\documentclass[aps,prl,reprint, superscriptaddress
]{revtex4-2}

\usepackage{graphicx}% Include figure files
\usepackage{dcolumn}% Align table columns on decimal point
\usepackage{bm}% bold math
\usepackage{hyperref}% add hypertext capabilities
\usepackage{color}
\usepackage{eufrak}

\begin{document}

\preprint{APS/123-QED}

\title{Reconstructing Hamiltonians from Correlations and Entanglement}

\author{J.~Alexander~Jacoby}
    \affiliation{Department of Physics, Brown University, Providence, Rhode Island 02912-1843, USA}
\author{J.~B.~Marston}%
    \affiliation{Department of Physics, Brown University, Providence, Rhode Island 02912-1843, USA}
    \affiliation{Brown Theoretical Physics Center, Brown University, Providence, Rhode Island 02912-1843, USA}

\date{\today}% It is always \today, today,
             %  but any date may be explicitly specified

\begin{abstract}
   We investigate two methods to reconstruct Hamiltonians of quantum matter, using a quantum spin chain to test them. The first method uses correlation functions and the second method uses entanglement spectra. The two methods are not specific to spin chains, and may find wider applicability to quantum matter that is sufficiently well characterized by experiment.
\end{abstract}

%\keywords{Suggested keywords}%Use showkeys class option if keyword
                              %display desired
\maketitle

\textit{Introduction}.~ Identification of Hamiltonians that describe the important degrees of freedom in quantum matter is the usual first step toward understanding their behavior. However, first-principles methods often fail for materials with strong correlations. Experiments measure thermodynamic properties and response functions but rarely point directly to a model. Quantum magnets that can be fully polarized in a strong field are a rare exception, as the magnon spectrum reveals the exchange interactions (see for instance \cite{Coldea}). New ways to build quantum many-body models from experimental data are needed. Here, we investigate two approaches to model reconstruction that rely upon the imprints of correlations and entanglement in quantum matter. We test the approaches using a well-understood quantum spin chain that has a rich phase diagram. We show that the Hamiltonian for the spin chain can be recovered from measurements of two-point correlation functions or, alternatively, from the spectra of entanglement.

 To compare the approach that uses correlations as the input with the entanglement-based method, we use a $SU(4)$ spin chain as the illustrative target system. We note that interest in $SU(n)$ quantum spin models is growing. For example, both experimental and theoretical developments suggest novel quantum phases may be realizable in $SU(n)$ symmetric ultra-cold atomic systems \cite{XVBS,Cold_Atom1,Cold_Atom2,Cold_Atom3,Cold_Atom4,Cold_Atom5,Cold_Atom6,Cold_Atom7} and such systems are also potential resource states for measurement based quantum computing \cite{Optical_Lattices,MBQC2,MBQC1}. Another example is the emergence of $SU(4)$ symmetry from mixed spin-orbital degrees of freedom \cite{Onufriev:1999fq,SU(4)QSOL4}. Some of these systems may be in a Quantum Spin-Orbital Liquid (QSOL) phase \cite{SU(4)QSOL0,SU(4)QSOL1,SU(4)QSOL2,SU(4)QSOL3,SU(4)QSOL5,SU(4)QSOL6}. Though some doubt has recently been cast upon $\alpha-{\rm Zr}{\rm Cl}_{3}$, a popular candidate material \cite{SU(4)QSOL7}, enlarged spin symmetries remain an enticing possibility.

\textit{Model}.~  $SU(2)$ AKLT chains \cite{AKLT} may be generalized to $SU(n)$ in several ways \cite{XVBS,GreiterGap1,GreiterGap2}. We use a model first introduced in \cite{XVBS}. We note that recent work \cite{SU(4)-QSL-1,SU(4)-QSL-2} suggests that such models could support spin-liquid phases in two dimensions. Two fermions in the fundamental representation of $SU(4)$ populate each site. The fermions on site $i$ are created and destroyed by $c^{\dagger\, \alpha}(i)$ and $c_{\alpha}(i)$, respectively, where $\alpha = 1, 2, 3, 4$ and repeated raised and lowered indices are summed over. $SU(4)$ spin operators are defined as $S^{\mu}_{\nu}(i) = c^{\dagger\,\mu}(i)c_{\nu}(i)-\frac{1}{2}$ and they satisfy the $SU(4)$ algebra, $\left[S^{\mu}_{\nu}(i),S^{\alpha}_{\beta}(j)\right] = (\delta^{\alpha}_{\nu} S_{\beta}^{\mu}+\delta^{\mu}_{\beta}S^{\alpha}_{\nu}) \delta_{ij}$. The Heisenberg spin-exchange operator, $\vec{S}_{i}\cdot\vec{S}_{i+1}$, generalizes to $SU(4)$ as $\mathcal{C}_{2}\left(i\right) = S^{\mu}_{\nu}\left(i\right)S^{\nu}_{\mu}\left(i+1\right)$, and $SU(4)$ singlets are formed using the four index antisymmetric tensor, $\epsilon_{\mu\nu\lambda\rho}$, which in addition to $\delta^{\mu}_{\nu}$, is an $SU(4)$-invariant tensor.

\begin{figure}[h!]
    \centering
    \includegraphics[width = 0.7\linewidth]{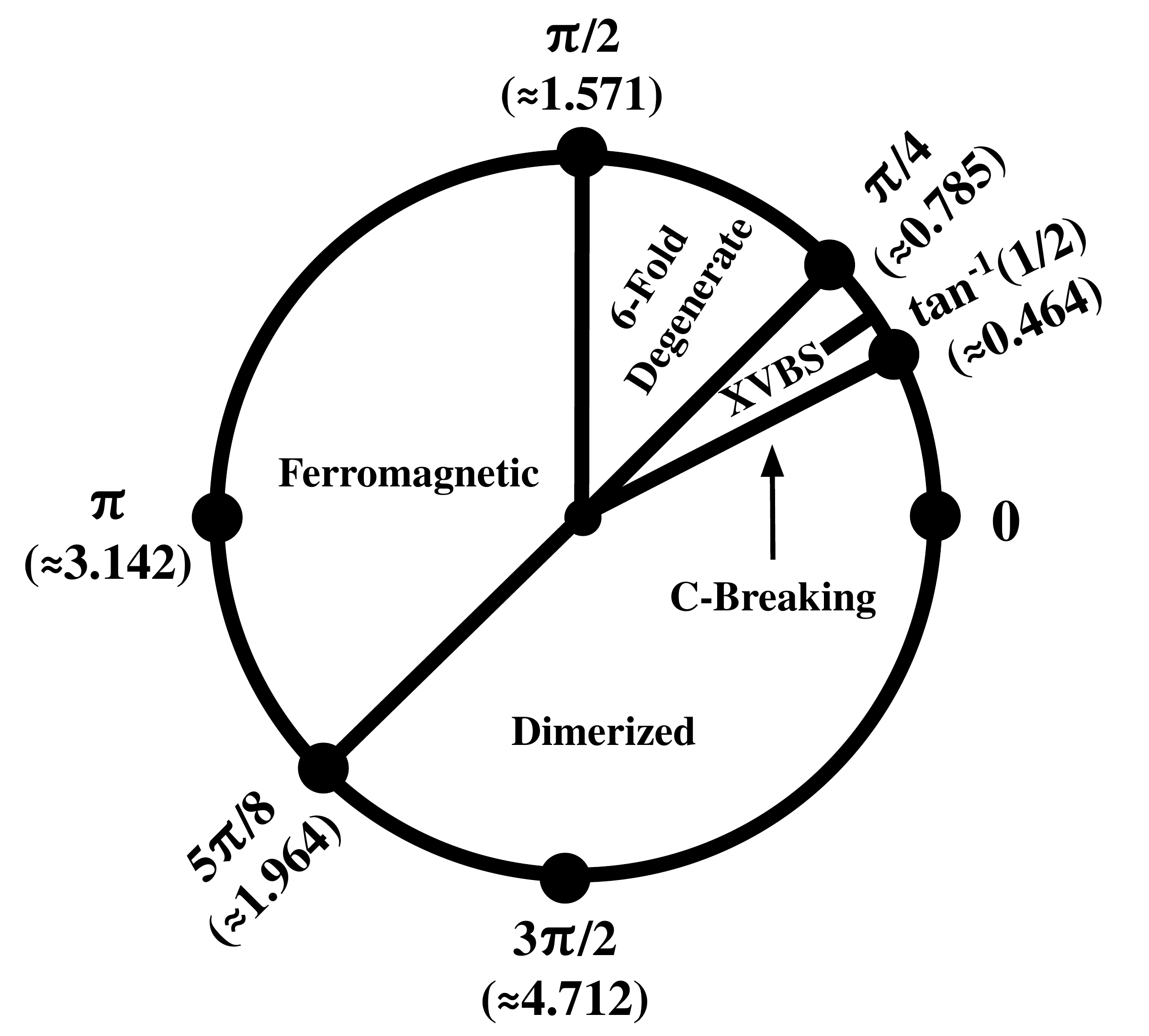}
    \caption{Phase diagram of the $SU(4)$ model with Hamiltonian, $H\left(\theta\right)$, given by Eqn. (\ref{eqn:modelHamiltonian}).}
    \label{fig:PD}
\end{figure}

Spin operators act on the antisymmetric, $\mathbf{6}$ dimensional representation of $SU(4)$ (there are six antisymmetric combinations of indices ranging over four values) and the Hamiltonian is parameterized by an angle $\theta$ (see Fig. \ref{fig:PD}),
\begin{eqnarray}
    H\left(\theta\right)&=&  = \sum_{i}^{L-1}H_{i}\left(\theta\right) \nonumber \\
    H_{i}\left(\theta\right) &=& \cos\left(\theta\right)\mathcal{C}_{2}(i)+\frac{\sin\left(\theta\right)}{4} \mathcal{C}_{2}^{2}(i)
    .
    \label{eqn:modelHamiltonian}
\end{eqnarray}
Note that these two operators, $\mathcal{C}_{2}\left(i\right)$ and $\mathcal{C}_{2}^{2}\left(i\right)$, are the only two $SU(4)$-invariant nearest neighbour couplings, and higher powers are mathematically redundant. At the point $\theta= \arctan\left(2/3\right)\approx 0.588$, $H_{i}\left(\theta\right)$ becomes a projector (up to an additive constant) onto the maximally-symmetric, $\mathbf{20}$ dimensional bond representation between sites $i$ and $i+1$. For a periodic chain, there are two degenerate ground states formed from $SU(4)$ singlet bonds that extend across three adjacent sites. This is called the eXtended Valence Bond Solid (XVBS) state and it is of the Matrix Product State (MPS) form. The two ground states break charge-conjugation and translational symmetries. To construct the ground state wavefunction of finite chains with boundary spins it is convenient to introduce the matrix operator:
\begin{equation}
    \Lambda^{\mu_{i}}_{\mu_{i+2}} = \epsilon_{\nu_{i}\mu_{i+1}\nu_{i+1}\mu_{i+2}}
    c^{\dagger\, \mu_{i}}c^{\dagger \, \nu_{i}}c^{\dagger\, \mu_{i+1}}c^{\dagger \, \nu_{i+1}}.
    \label{eqn:xvbsmps}
\end{equation}
For clarity here and in what follows, site indices on the fermion creation and annihilation operators are implied by the subscripts on the spin indices.
%From Eq. (\ref{eqn:xvbsmps}) we see that the singlets must stretch out across two sites, as a single matrix, $\Lambda^{i}_{i+2}$, corresponds to a singlet and the two indices are associated to two sites, $i$ and $i+1$, which are separated by a another full site.
The ground state may then be written as
\begin{equation}
    \left|\Psi \right> =\mathcal{B}_{\mu_{1}\nu_{L}}c^{\dagger\,\mu_{L}}c^{\dagger\,\nu_{L}} \prod_{i=1}^{(L-1)/2}\Lambda^{\mu_{2i-1}}_{\mu_{2i+1}}\left|0\right>,
    \label{eqn:MPS}
\end{equation}
where the complex matrix, $\mathcal{B}$, sets the boundary condition on the two sites at the end of the chain. Over the slice of $\theta$-space shown in Fig. \ref{fig:PD} that contains the XVBS state, charge-conjugational symmetry is spontaneously broken for infinite chains and chains on periodic lattices (those without a boundary). For finite chains, charge-conjugational symmetry is explicitly broken by the boundary modes.

\textit{Entanglement}.~ Consider a many-body wavefunction, $\vert \Psi \rangle $, and the binary partitioning of its degrees of freedom into two sets, ${\rm A}$ and ${\rm B}$. $\vert \Psi \rangle $ can be decomposed as
\begin{equation}
    \left|\Psi\right> = \sum_{i} \lambda_{i}\left|\psi^{{\rm A}}_{i}\right>\otimes \left|\psi^{{\rm B}}_{i}\right> \quad \mathcal{S}_{vN} = -\sum_{i}{\rm Log}\left[\lambda_{i}\right],
    \label{eqn:Schmidt_decomp}
\end{equation}
where $\mathcal{S}_{vN}$ is called the von Neumann entropy and the $\lambda_{i}$ are called the Schmidt values. $\mathcal{S}_{vN}$ may also be defined in terms of a reduced density matrix, $\rho_{{\rm A}/{\rm B}} = {\rm Tr}_{{\rm B}/{\rm A}}\left[\left|\Psi\right>\left<\Psi\right|\right]$ with $\mathcal{S}_{vN} = - {\rm Tr}\left[\rho_{{\rm A}/{\rm B}} {\rm Log}\left[\rho_{{\rm A}/{\rm B}}\right]\right]$. A notable feature of $S_{vN}$ is ${\rm A}-{\rm B}$ inversion symmetry: the entanglement between subsytems cannot depend upon which one is traced out. $\mathcal{S}_{vN}$ characterizes the amount of information about ${\rm A}$ contained in ${\rm B}$ and vice versa.

\begin{figure}[h]
    \centering
    \includegraphics[width = \linewidth]{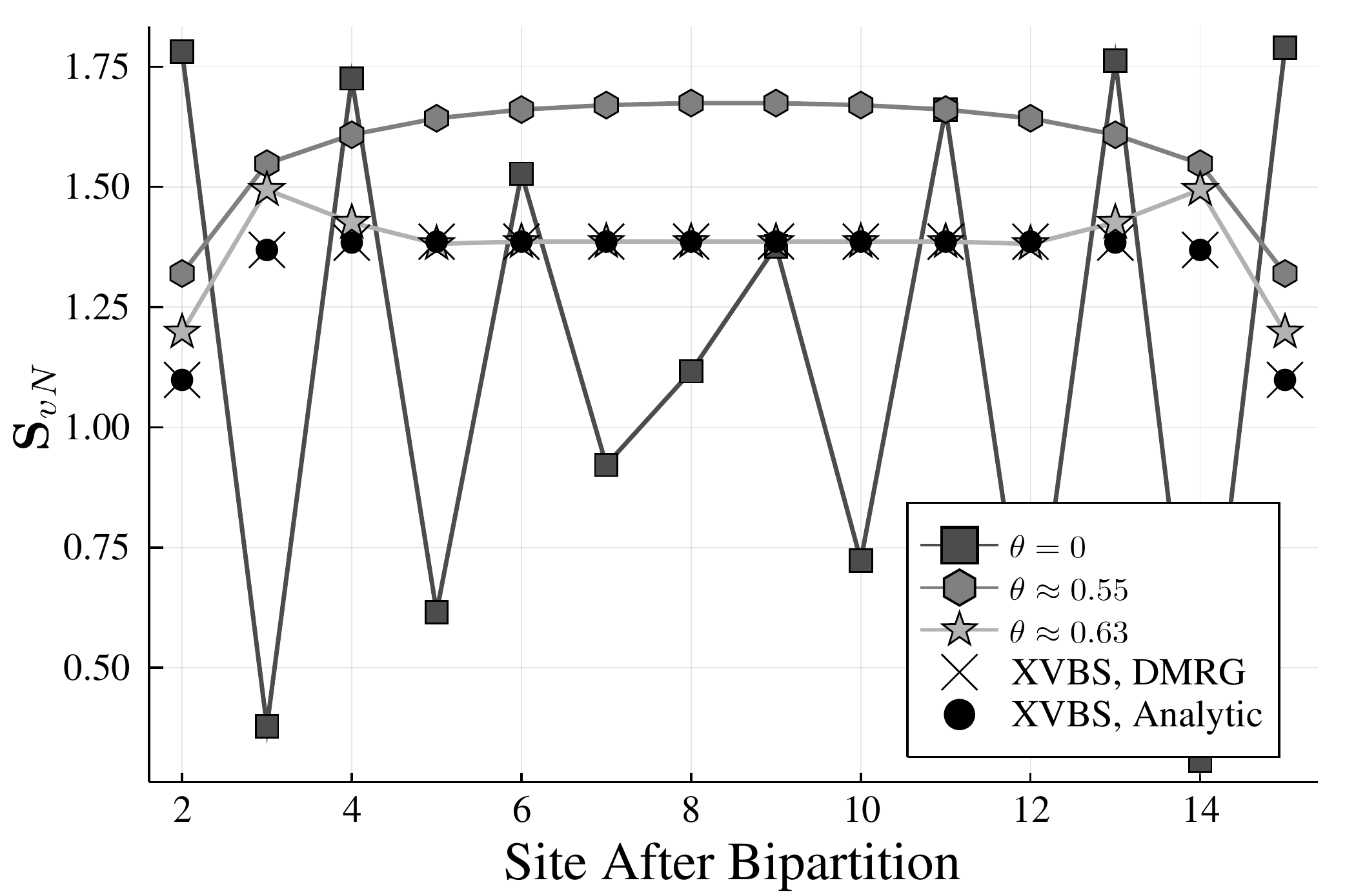}
    \caption{The entanglement entropy of a 15 site chain evaluated by DMRG at various $\theta$ and analytically at the XVBS point. The analytically calculated entropy at the XVBS point agrees with DMRG. As $\theta$ increases from the dimerized to the C-breaking phase, the sawtooth pattern melts and the edges inflect down to accommodate the fractionalized edge modes. Away from the XVBS point the block Hilbert space size, $M$, is set to $40$.}
    \label{fig:entanglement}
\end{figure}

Entanglement is well studied for the exact VBS states \cite{AKLT_Entanglement1,AKLT_Entanglement2,AKLT_Entanglement3,VBS_Entanglement,AKLT_Entanglement4} and, after generalization to the XVBS states, can be used to check Density Matrix Renormalization Group (DMRG) \cite{DMRG_White} calculations. We find the entanglement entropy for an XVBS chain of length $2N+1$ with a bipartition at site $2n+1$ to be:
\begin{equation}
    \mathcal{S}_{vN}= \frac{\lambda_{1}}{2\lambda_{3}}{\rm Log}\left(\frac{\lambda_{1}}{2\lambda_{3}}\right)-\frac{\lambda_{2}}{2\lambda_{3}}{\rm Log}\left(\frac{\lambda_{2}}{6\lambda_{3}}\right),
    \label{eqn:ee-exact-sym}
\end{equation}
where $\lambda_{1}=\chi_{n}+\chi_{N-n}-2\chi_{n}\chi_{N-n}$, $\lambda_{2}=4-5\chi_{n}-5\chi_{N-n}+6\chi_{n}\chi_{N-n}$, and $\lambda_{3}= 4\chi_{n}\chi_{N-n}-3\chi_{n}-3\chi_{N}+2$. $\chi_{n}$ is defined recursively by the relations $\chi_{0}=1$, $\chi_{1}=3$, and $\chi_{n+1}= 3\left(3\chi_{n}-2\right)$. Here the boundary condition on the chain ends are chosen to be symmetric (a state in the $\mathbf{10}$ representation). For example, we could select $\mathcal{B}_{1 1}=1$ with all other matrix elements set to zero. Comparison is made to DMRG using the DMRjulia implementation \cite{Baker_2021} in Fig. \ref{fig:entanglement} with precise agreement for block Hilbert space sizes $M \geq 6$ as expected for an exact matrix product state.

\textit{Reconstruction from correlation functions.}~  The first reconstruction method \cite{Qi} makes use of correlation functions of $\Psi$. General correlation functions include operators acting at arbitrary separations. Locality may be imposed by restricting to range-k operators which act at separations of at most $k$ sites. Various symmetries may also be assumed, reducing the number of allowed operators. Define a vector of such operators, $\vec{\mathcal{O}}$, and an ansatz Hamiltonian, $H\left(\vec{w}\right)=\vec{w}\cdot \vec{\mathcal{O}}$. We assume for simplicity that $\vec{w}\left(\theta\right) = ( \cos\left(\theta\right),\frac{\sin{\theta}}{4})$, $\vec{\mathcal{O}} = \left(\sum_{i}\mathcal{C}_{2}(i),~ \sum_{i} \mathcal{C}_{2}^2(i)\right)$, and we identify $H\left(\theta\right) = H\left(\vec{w}\left(\theta\right)\right)$. We define the correlation matrix of state $\Psi$ as
\begin{equation}
    \mathcal{M}_{mn}^{\Psi} = \frac{1}{2} \left<\mathcal{O}_{m}\mathcal{O}_{n}\right>_{\Psi}-\left<\mathcal{O}_{m}\right>_{\Psi}\left<\mathcal{O}_{n}\right>_{\Psi}
    \label{eqn:cormat}
\end{equation}
where $m, n = 1, 2$. When $\mathcal{M}^{\Psi}\vec{w}=0$, we find that $\Psi$ is an eigenstate of $H\left(\vec{w}\right)$ since an eigenstate of an operator has no variance in eigenvalue. Thus, the correlation matrix is guaranteed to have at least one zero eigenvalue, corresponding to a Hamiltonian (up to an overall constant factor) with $\Psi$ as an eigenstate.

\begin{figure}[h]
    \centering
    \includegraphics[width = \linewidth]{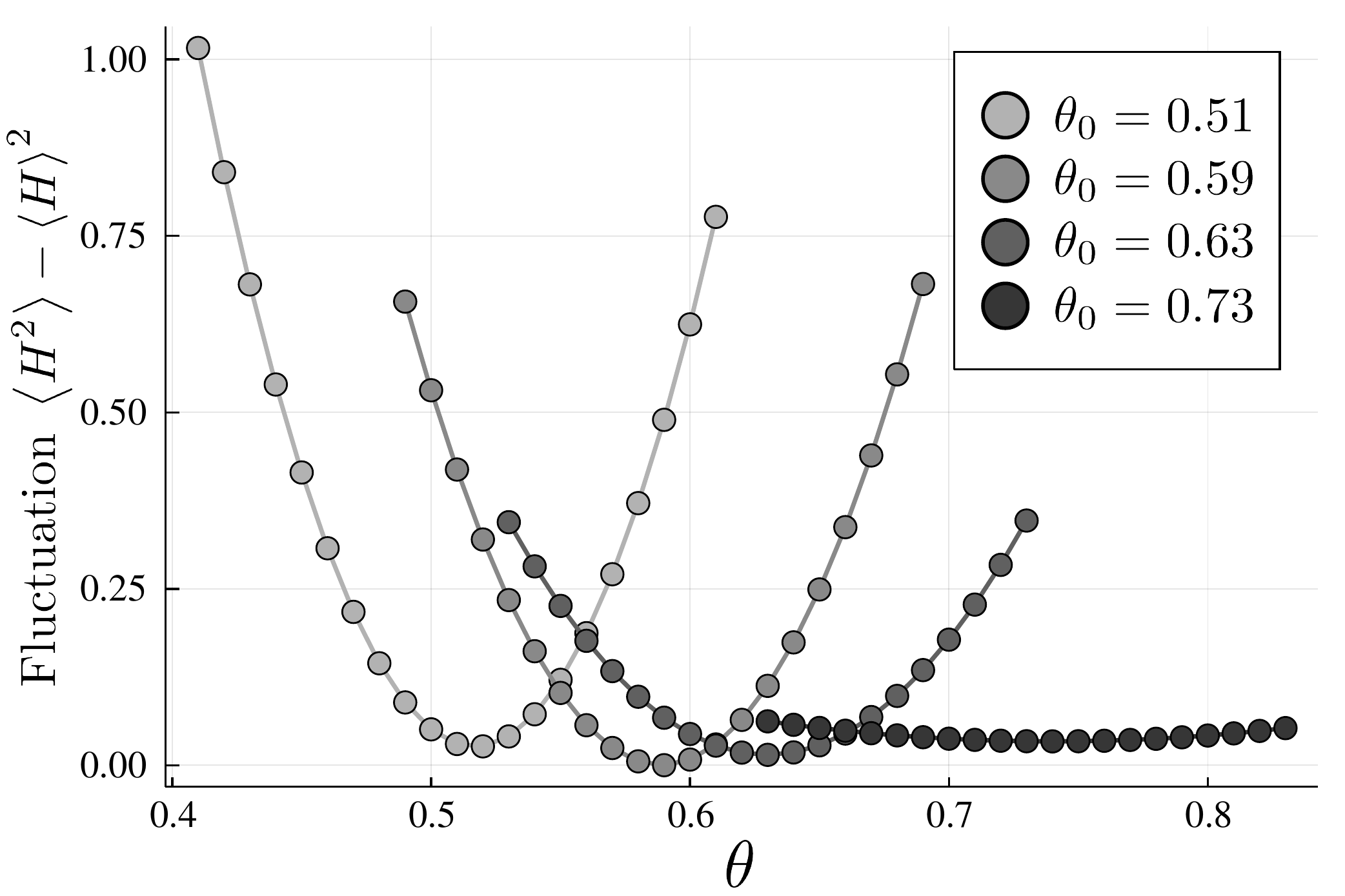}
    \caption{Fluctutation or variance of $H\left(\theta\right)$, evaluated in the groundstate of $H\left(\theta_{0}\right)$ with chain length $L=10$ with block Hilbert space size $M = 40$. All curves have well defined minima and are convex on the region considered.}
    \label{fig:fluct}
\end{figure}
A key feature of the method presented in \cite{Qi} is the uniqueness of a reconstructed Hamiltonian. Beyond selecting \textit{a} Hamiltonian with $\Psi$ as an eigenstate, it is unlikely that $\mathcal{M}^{\Psi}$ has a second zero eigenvalue (and therefore a second candidate Hamiltonian) if at least one point in phase space has a unique Hamiltonian, as is argued in \cite{Qi}. For example, consider the variance, or fluctuation, of $H\left(\theta\right)$ in the ground state of $H\left(\theta_{0}\right)$ (found using DMRG) at various $\theta_{0}$, as displayed in Fig. \ref{fig:fluct}. The fluctuation of $H\left(\theta\right)$ is convex in $\theta$ in the region examined, assuring the existence of a minimum and therefore a unique reconstructed Hamiltonian. The minimum is lowest at the XVBS point ($\theta\approx 0.59$), where the ground state is an MPS.

\begin{figure}[h]
    \centering
    \includegraphics[width = \linewidth]{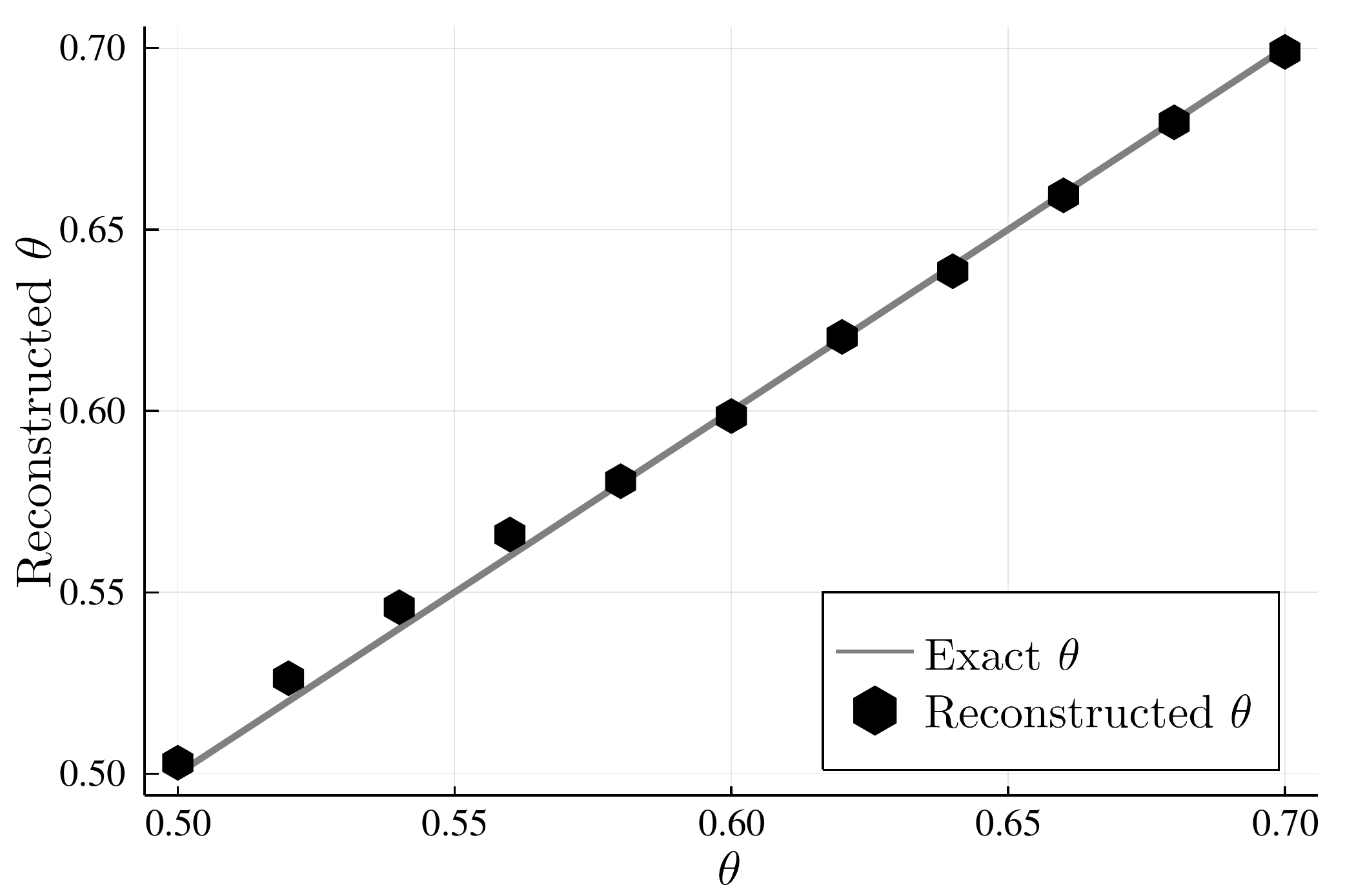}
    \caption{Correlation Matrix Reconstructed Values of $\theta$. Correlation functions were computed with the same wavefunctions as in Fig. \ref{fig:fluct} with DMRG.}
    \label{fig:fluct_recon}
\end{figure}

In Fig. \ref{fig:fluct_recon} we demonstrate the recovery of $\theta$ at the zero eigenvalue of the correlation matrix defined by Eq. (\ref{eqn:cormat}). Error in the reconstructed $\sin(\theta)$ is on the order of the fluctuation minima of Fig. \ref{fig:fluct}, showing that the correlation matrix does not add much if any additional noise to the input data in this case.
This is a way to reconstruct a Hamiltonian from $\Psi$ or a full set of correlation functions. While correlation functions with multiple spin operators, also called multipolar operators, are traditionally very challenging to measure, progress has recently been made \cite{HigherCorrelations1,HigherCorrelations2,HigherCorrelations3}. In the spirit of \cite{HigherCorrelations1} a constraint is that we consider only operators which couple directly to the experimental probe used (in \cite{HigherCorrelations1}, this is developed within the formalism of Resonant Inelastic X-ray Scattering, or RIXS, for magnetic insulators). A potential impediment for the full reconstruction of a Hamiltonian is that not all microscopically consequential operators are likely to couple to conventional experimental probes.

\textit{Reconstruction from entanglement}.~ An entanglement spectrum can also be used to reconstruct the Hamiltonian. This makes sense intuitively, as reduced density matrices are generically recoverable from correlation functions \cite{DM_COR}. The entanglement spectrum methodology is rooted in the understanding of the reduced density matrix as thermal in a sense we will make precise below. The existence of such a thermal form of the reduced density matrix, with some more technical stipulations, is equivalent to the satisfaction of the Eigenstate Thermalization Hypothesis (ETH) and holds beyond the ground state (see \cite{Grover}). We can interpret an entangled state, $\left|\Psi\right>$, for system ${\rm AB}$, restricted to ${\rm A}$ (which is assumed to hold fewer degrees of freedom than ${\rm B}$) as a statistical ensemble. This identification is made via the Schmidt decomposition, Eqn. \ref{eqn:Schmidt_decomp}. That is, we can identify the Schmidt values with the Boltzmann weights in the sense of $e^{-\beta E_{i}} \sim \lambda_{i}$ for microstates $\left\{\left|\psi^{{\rm A}}_{i}\right> \right\}$. This choice is motivated by the desire for equality between expectation values (of operators restricted to ${\rm A}$) taken with a reduced density matrix on ${\rm A}$ and via the statistical operator, $e^{-\beta H}$. We want $\frac{1}{\mathcal{Z}}{\rm Tr}\left[\left.\mathcal{O}\right|_{{\rm A}}e^{-\beta H}\right] ={\rm Tr}\left[\left.\mathcal{O}\right|_{{\rm A}}\rho_{{\rm A}}\right] $, where $\mathcal{Z}= {\rm Tr}\left[e^{-\beta H}\right]$ is the standard partition function. In general, we assume the reduced density matrix possesses the form $\rho_{{\rm A}} = e^{-\beta H_{{\rm A}}}$ with $\left\{\lambda_{i}\right\}$ the same as in Eqn. (\ref{eqn:Schmidt_decomp}), $\beta$ the  entanglement temperature, and $H_{{\rm A}}$ the  entanglement Hamiltonian. The entanglement Hamiltonian must be a sum of local operators for this identification to make sense since the energy of the microstates of ${\rm A}$ will only depend upon degrees of freedom within ${\rm A}$ and near the bipartition interface between ${\rm A}$ and ${\rm B}$.

The Bisognano-Wichmann (BW) theorem in algebraic quantum field theory sets an explicit form for $H_{{\rm A}}$, which we refer to as the BW ansatz. Recent work \cite{BW1,BW2,BW3,BW4}, has adapted this ansatz to a lattice (for less than half space bipartitions):
\begin{equation}
    H_{{\rm A}}^{{\rm BW}} = \sum_{i}\left|x_{i}\right| H_{i} \qquad \sigma^{{\rm BW}} = \exp\left[-\beta H_{{\rm A}}^{{\rm BW}}+c\right]
    \label{eqn:BW_ansatz}
\end{equation}

\begin{figure}[h]
    \centering
    \includegraphics[width = \linewidth]{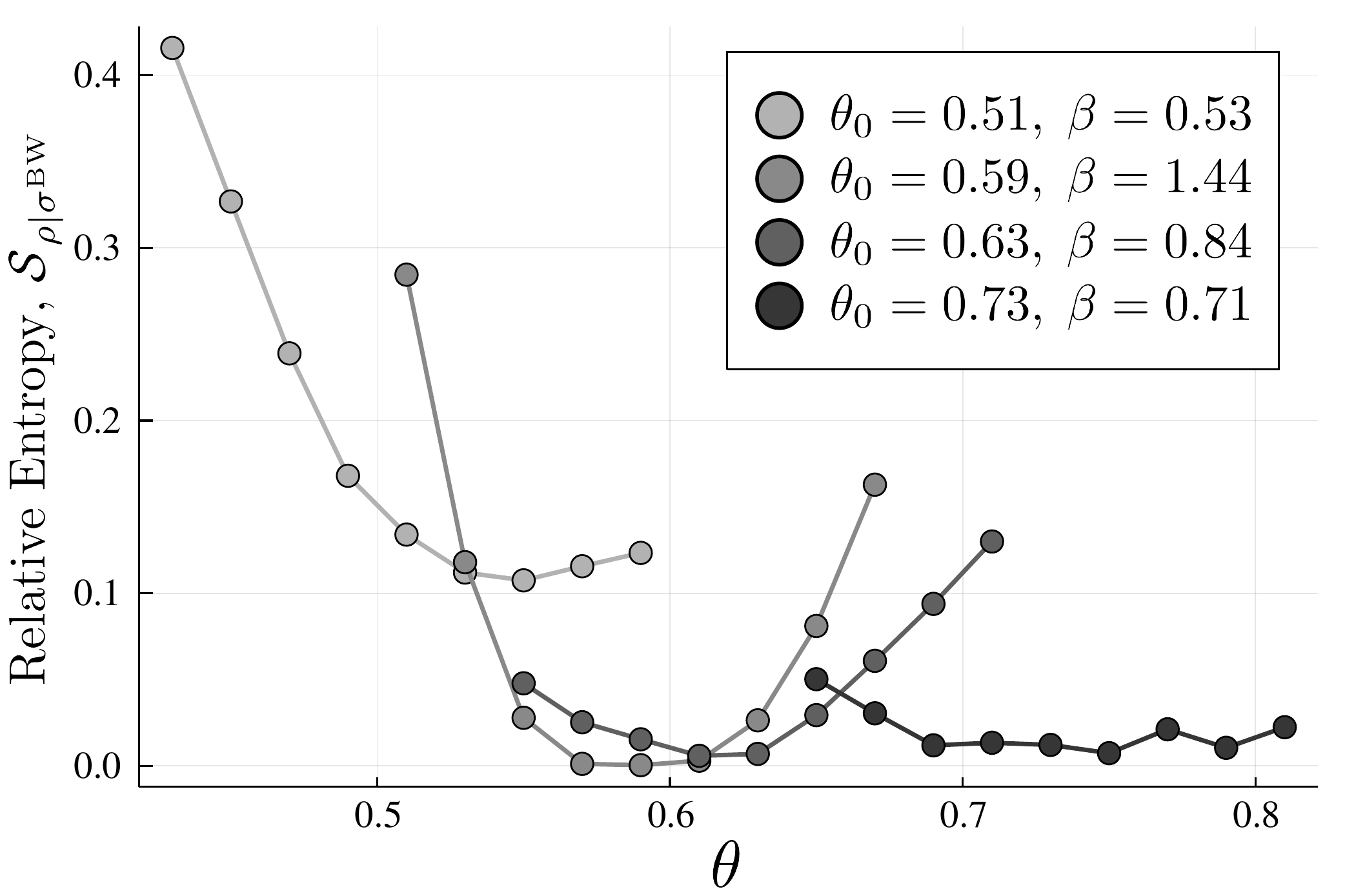}
    \caption{Relative entropy of the BW ansatz and DMRG-calculated reduced density matrix for $7$ site subsystem of a $15$ site chain. The relative entropy between reconstructed and original density matrices is defined as $\mathcal{S}_{\left.\rho \right| \sigma^{\rm BW}} = \rho\ {\rm Log}\left[\rho\right]-\rho\ {\rm Log}\left[\sigma^{\rm BW}\right]$.}
    \label{fig:trdist}
\end{figure}

Eq. (\ref{eqn:BW_ansatz}) merits a more physical description. We consider only short-range entangled, gapped phases here. Correlations between two parts of a system require that one part raises its energy to lower the other part's energy and vice versa by superposition, inducing entanglement when the correlation is cleaved by a bipartition. Correlations between the subsystems decay exponentially with distance. The factor of $\left|x_{i}\right|$ in the summand filters the relevant degrees of freedom, which sit near the boundary and are likeliest to be entangled across it, by increasing as we move away from the bipartition, suppressing far-away degrees of freedom via the negative exponential. In the XVBS state, for example, the relevant degrees of freedom which sit near the boundary are simple: they are the boundary indices of the MPS at the bipartition interface. Thus, $\beta$ can be interpreted as measure of how quickly entanglement decays in the distance from the bipartition. Alternatively, the entanglement temperature is high near the bipartition and lowers away from it. Typically, $\beta$ is set to $2\pi/c$ where $c$ is the relativistic velocity; in our calculations, $\beta$ is fitted numerically since no such relativistic velocity is known. Away from the bipartition boundary, $\sigma^{BW}$ acts just like the Boltzmann weight $\rho = e^{-\beta H}$. Application of the methodology to the $SU(4)$ chain is shown in Fig. \ref{fig:trdist}. To reconstruct $H$ the relative entropy between the observed entanglement spectrum and $\sigma^{\rm BW}$, $\mathcal{S}_{\left.\rho \right| \sigma^{\rm BW}} = \rho\ {\rm Log}\left[\rho\right]-\rho\ {\rm Log}\left[\sigma^{\rm BW}\right]$, is minimized in $\vec{w}$. There is an obvious resemblance to Fig. \ref{fig:fluct}.

There are several limitations of this approach. Equality is only satisfied in the thermodynamic limit and excitations should be described by a relativistic field theory. The approach is also not as robust as the approach based upon correlations. Perhaps the main limitation is that the determination of entanglement spectra requires quantum state tomography, which is the challenging reconstruction of a quantum state by a series of measurements \cite{ZollerEntanglement}. Despite all these constraints, the Bisognano-Wichmann ansatz technique provides an alternative to the correlation matrix technique of the previous section in this case. Nonetheless, the relationship between entanglement and energy spectra is not fully understood and must be treated with caution (see \cite{ES_Sondhi}).

\textit{Conclusion}.~ In this Letter, we explored the correlation matrix \cite{Qi}, and Bisognano-Wichmann ansatz \cite{BW1,BW2,BW3,BW4}, techniques to reconstruct Hamiltonians. The first method requires a set of correlation functions and the second method requires entanglement spectra as experimental inputs. While these inputs are not readily accessible, both methods are computationally accessible. Further investigation with a variety of models is necessary to establish the efficacy of the two techniques. For the correlation matrix method, determining which correlation functions can be neglected would improve efficiency in systems with fewer symmetries. Constraints beyond the symmetry group, locality, and Hermicity would be equally useful. For the BW technique, a more systematic method of calculating $\beta$ should be possible in systems without obvious mappings to field theories, though it is not clear whether the BW technique will be effective in these systems. The relationship between spontaneous symmetry breaking, loss of Lorentz invariance, and the BW ansatz's effectiveness is also currently unclear. Finally, the BW approach's robustness against noisy data must be determined before it is ready to be applied.

\textit{Acknowledgements}.~ We are grateful for discussions with V. Mitrovi\'c and K. Plumb. J.~A.~J. would like to extend a special thanks to S. Carr. This work was supported in part by U.S. National Science Foundation Grants No. OIA-1921199 and No. OMA-1936221. J.~A.~J. was also partly supported by a Karen T. Romer Undergraduate Teaching and Research Award (UTRA). This research was conducted using resources at the Center for Computation and Visualization at Brown University.

% The \nocite command causes all entries in a bibliography to be printed out
% whether or not they are actually referenced in the text. This is appropriate
% for the sample file to show the different styles of references, but authors
% most likely will not want to use it.
%\nocite{*}

\bibliography{ms}% Produces the bibliography via BibTeX.

\end{document}